\documentclass[12pt]{article}
\newcommand{\tr}{\mbox{Tr}}
\newcommand{\eqn}[1]{(\ref{#1})}
\newcommand{\del}{\partial}
\newcommand{\zed}{{\bb Z}} 
\newcommand{\id}{{\bb I}} 

\font\mybb=msbm10 at 12pt
\def\bb#1{\hbox{\mybb#1}}

\def\slash{\!\!\!\!/}

\def\be{\begin{equation}}
\def\ee{\end{equation}}
\def\bea{\begin{eqnarray}}
\def\eea{\end{eqnarray}}
\def\bd{\begin{displaymath}}
\def\ed{\end{displaymath}}

\makeatletter
\newdimen\normalarrayskip              
\newdimen\minarrayskip                 
\normalarrayskip\baselineskip
\minarrayskip\jot
\newif\ifold             \oldtrue            
\def\arraymode{\ifold\relax\else\displaystyle\fi} 
\def\@arrayskip{\ifold\baselineskip\z@\lineskip\z@
     \else
     \baselineskip\minarrayskip\lineskip2\minarrayskip\fi}
\def\@arrayclassz{\ifcase \@lastchclass \@acolampacol \or
\@ampacol \or \or \or \@addamp \or
   \@acolampacol \or \@firstampfalse \@acol \fi
\edef\@preamble{\@preamble
  \ifcase \@chnum
     \hfil$\relax\arraymode\@sharp$\hfil
     \or $\relax\arraymode\@sharp$\hfil
     \or \hfil$\relax\arraymode\@sharp$\fi}}
\def\@array[#1]#2{\setbox\@arstrutbox=\hbox{\vrule
     height\arraystretch \ht\strutbox
     depth\arraystretch \dp\strutbox
     width\z@}\@mkpream{#2}\edef\@preamble{\halign \noexpand\@halignto
\bgroup \tabskip\z@ \@arstrut \@preamble \tabskip\z@ \cr}%
\let\@startpbox\@@startpbox \let\@endpbox\@@endpbox
  \if #1t\vtop \else \if#1b\vbox \else \vcenter \fi\fi
  \bgroup \let\par\relax
  \let\@sharp##\let\protect\relax
  \@arrayskip\@preamble}
\makeatother

\newlength{\extraspace}
\newlength{\extraspaces}
\begin{document}
\title{{\footnotesize
{\hfill $\begin{array}{r}
                \mbox{DSF 29/2000} \\
                \mbox{hep-th/0009180}
\end{array}$}}\\ $~$ \\
{\large \bf Another Alternative to Compactification: \\ Noncommutative
Geometry and Randall-Sundrum Models}}
\author{Fedele Lizzi\thanks{lizzi@na.infn.it},
Gianpiero Mangano\thanks{mangano@na.infn.it}, and Gennaro
Miele\thanks{miele@na.infn.it}\\ \\ Dipartimento di Scienze Fisiche,
Universit\`{a} di Napoli {\sl Federico II},\\ and INFN, Sezione di Napoli,
I-80126 Napoli, Italy}
\date{}
\maketitle
\begin{abstract}
We observe that the main feature of the Randall--Sundrum model, used to
solve the hierarchy problem, is already present in a class of Yang--Mills
plus gravity theories inspired by noncommutative geometry. Strikingly the
same expression for the Higgs potential is found in two models which have
no apparent connection. Some speculations concerning the possible
relationships are given.
\end{abstract}
\newpage
The difference between the electroweak and the Planck scales is sixteen
orders of magnitude, any good model of fundamental interactions at high
energies should be able to account for such a huge hierarchy. It would be
desirable to have the two scales emerging from a common mechanism,
proposals for this have been put forward in the framework of models with
large extra dimensions \cite{AHDD,RS}. These dimensions are not visible
because fermions, and matter fields in general, are constrained to live on
a four dimensional slice. String theory, and branes in particular, may, to
some extent, account for this phenomenon.

The success of the Randall--Sundrum model \cite{RS} (RS) is due to the
efficient and suggestive way the hierarchy problem is understood. In the
simplest case of just one extra dimension, and two branes, with a suitable
ansatz for the five dimensional metric, the electroweak scale is obtained
in terms of the Planck mass as $m_{ew}=m_{Pl}
\exp(-\pi k r_c)$, where $r_c$ is the size of the extra dimension, $k$ a
constant with $k r_c\sim 50$. Therefore no fine tuning is necessary to get
the correct hierarchy.

In this letter we argue that this goal can be also achieved with a model
\cite{CL} which draws its inspiration and techniques from Noncommutative
Geometry \cite{Connes}, although the geometry described is still that of an
ordinary manifold. In this model, the physical space is assumed to be a
two-sheeted manifold, ${\cal M} {\times}
\zed_2$, with ${\cal M}$ the ordinary four dimensional spacetime. Left and
right-handed fermions live on the two different sheets, which are coupled
by a scalar field representing the component of the connection in the
``discrete" direction, which turns out to be the Higgs field. In this
framework the natural mass scale for this field should be the Planck one,
being related to the structure of spacetime, but at the same time it must
reproduce the known phenomenology at the electroweak scale. The ratio of
these two fundamental scales is actually proportional to the distance
between the two sheets measured in Planck units. The hierarchy problem is
then solved provided there is a mechanism able to stabilize this distance,
as in the RS model. The novelty is that in the model here presented this
distance itself naturally corresponds to a dynamical field, being related
to the component of gravity in the discrete direction. In \cite{LMMS} we
have explored the possibility that this field is in fact the one that
drives inflation.

Although the Connes-Lott model with gravity \cite{CFF} and the RS
constructions are quite different, as the former was in fact developed
before and independently, they share the characteristic of treating the
matter fields in a different way than the gravity fields, with the latter
living in a wider setting. The analogy between the two branes and the two
sheets of spacetime is probably more coincidental, as there are version of
the RS model with a different number of branes, while on the other side
slightly different setting of Noncommutative Geometry also lead to
exponential corrections of the discrete component of gravity (see e.g.\
example 8.3.2 of \cite{Madore}). A further discussion on these analogies is
at the end of the paper.

We will start with a very brief review of the Connes--Lott model \cite{CL},
and its version with gravitational interaction introduced by Chamseddine,
Felder and Fr\"{o}hlich \cite{CFF} (see also \cite{ncggrav}). In this
framework, the effective Higgs potential has the very same expression found
in \cite{RS}, with the exponential correction which solves the hierarchy
problem. We will be very sketchy in our description, and refer to the
original literature, or the review \cite{VG-BM} for more details. At the
end we then make an effort to further tie the two models.

The idea behind noncommutative geometry is to describe geometrical spaces
using the algebras of fields rather than their punctual properties. This
has become popular in the last year \cite{SW} with the use of deformations
of the function algebras on a manifold, which could describe situations
where the coordinates do not commute, so it would be impossible to describe
spacetime as an usual set of points. It is however useful to take this
point of view even to describe ordinary spaces. Connes and Lott \cite{CL}
showed that considering the algebra of functions on a two-sheeted
spacetime\footnote{It may be useful to think of this space as a sort of
discrete Kaluza--Klein with the internal space composed of a pair of
points.}, the Higgs field, together with its quartic potential emerges
naturally. The key ingredient of the theory is the {\it spectral triple}
$({\cal A},{\cal H},D)$, with ${\cal A}$ an {\it algebra}, which in the
case we are considering is the algebra of functions on ${\cal M} {\times}
\zed_2$, $D$ an {\it operator} which generalizes the Dirac operator, and
${\cal H}$ the {\it Hilbert space} of physical fermions, in our example
$L^2({\cal M}, S)_R \oplus L^2({\cal M}, S)_L$, the direct sum of Hilbert
spaces of square integrable sections of left and right handed spinor
bundles, onto which both the algebra and $D$ act.

A generic element $a$ of the algebra of (continuous) functions on ${\cal M}
{\times} \zed_2$ can be usefully represented by a diagonal $2{\times} 2$ matrix:
\be
a=\left(\begin{array}{cc} a^L(x)& 0 \\ 0 &a^R(x)\end{array}\right)~~~,
\label{alg2points}
\ee
where $x \in {\cal M}$, and $a^L(x)$ and $a^R(x)$ are complex number or
matrix valued functions. In the first case, which we mostly use as more
explanatory, what is obtained is a $U(1)$ gauge theory with a complex Higgs
field.  Gauge theories with larger groups require the $a^{L,R}$ to be
matrices. This algebra acts on the Hilbert space of fermions ${\cal H}$,
which is naturally split into subspaces corresponding to left and right
chiralities. Every fermion can be seen as a column vector on which $a$
acts. In case of $n_g$ fermion generations the algebra is  represented as
$2n_g{\times} 2n_g$ matrix by tensor multiplying
\eqn{alg2points} by the identity matrix $\id_{n_g}$.

The generalization of the Dirac operator will act as a sum of two pieces: a
diagonal term, which is the usual derivative operator, and an extra term
which has a non-trivial behaviour along the discrete direction, acting as a
finite difference operator
\be
D=\left(\begin{array}{cc} \del\,\slash\, \otimes \id_{n_g}& \gamma_5 m
\otimes K
\\
\gamma_5 m \otimes  K^\dagger &\del\,\slash\, \otimes \id_{n_g} \end{array}\right)
~~~,\label{Diracgauge}
\ee
with $m$ a mass scale and $K$ a $n_g{\times} n_g$ matrix. The largest eigenvalue
of $K$ represents the inverse of the distance between the sheets, in unit
of $m^{-1}$.

In the usual case, gauge connections can be obtained in terms of elements
of the algebra by the generic expression
$A\slash\equiv\sum_ia_i(\del\,\slash b_i)$. Formally this means that
one-forms are represented via the commutator with the Dirac
operator\footnote{There are several technical complications related with
the appearance of the so-called junk-forms, forms which are zero, but whose
differential does not vanish. We refer to the literature for a proper
treatment of those problems.}. In the setting of the two sheeted manifold
this means that potentials are matrices as well:
\be
A=\sum_{j} a_{j} db_{j} = \sum_{j} a_{j} [D,b_{j}] =
\left( \begin{array}{cc} {A\!\!\!/}^{L} \otimes \id_{n_g} & \gamma_{5} \Phi
\otimes K\\
\gamma_{5} \Phi^{\dag} \otimes K^\dagger  & {A\!\!\!/}^{R} \otimes \id_{n_g}
\end{array} \right)~~~, \label{gaugepot}
\ee
where $\Phi =  m \sum_{j} a^{L}_{j} (K b^{R}_{j} - b^{L}_{j} K)$ and is
proportional to the inverse of the distance between the sheets. With the
potential one form $A$ it is possible to define a covariant Dirac operator
$D_A=D+A$, which will be a matrix as well, and to calculate the curvature
and  the bosonic part of the gauge action. The fermionic part in the
formalism may be cast in the form $\tr\,\bar\psi D_A \psi$. The gauge
action can be now evaluated still assuming the customary expression
\be
S_{A}= \int_{{\cal M}} d^4x ~\tr\left(d A + A^2\right)^2~~~,
\label{SA}
\ee
where now the exterior derivative is again defined making use of the
generalized Dirac operator, $d A\equiv\sum_{j} [D,a_{j}] [D,b_{j}]$, where
a quotient over junk forms is understood. Notice that the notion of
integration on ${\cal M} {\times} \zed_2$ is simply a four dimensional continuous
integral and a trace over the discrete $\zed_2$ degree of freedom. It is
remarkable to observe that this action is exactly the one of a $U(1)$ gauge
theory with spontaneous breaking of the symmetry, with $\Phi$ the Higgs
field \cite{CL}. We remind that $\Phi$ is a component of the gauge
potential $A$ in \eqn{gaugepot}. The interpretation is now clear. On the
two sheets live the two potentials ${A\!\!\!/}^{L(R)}$, while the off
diagonal $\Phi$ connects the two sheets. It is then possible to see the
Higgs field as the component in the discrete direction of the intermediate
``vector'' boson. The model can be rendered also more phenomenologically
valid with more realistic algebras, and also phenomenological predictions
for the standard model can be obtained, see for example \cite{Connesreal,
Marseille}.

Our next task is to add gravity, and this can be done by defining an
appropriate gravitational connection \cite{CFF,Landi}. In terms of this the
(euclidean) Dirac operator \eqn{Diracgauge} becomes
\be
D=\left(\begin{array}{cc} \nabla\slash \otimes \id_{n_g}& \gamma_5\phi
\otimes K \\
\gamma_5\phi^\dagger\otimes K^\dagger
&\nabla\slash\otimes \id_{n_g}\end{array}\right) ~~~,\label{Diracgrav}
\ee
with $\nabla\slash=\gamma^ae^\mu_a(\del_\mu+\omega_\mu)$ the gravitational
covariant derivative and $\omega_\mu$ the spin connection. Adding the
$\omega_\mu$ in the diagonal entries does not alter the gauge connection,
which is obtained with commutators. The off-diagonal terms of the Dirac
operator are the discrete components of the spin connection on the
two--sheeted manifold.

A generic (hermitean) connection $\Omega$ on the space of one--forms will
be a matrix
\begin{equation}
{\Omega_M}^N = \left(\begin{array}{cc}
             \gamma^\mu~ {\Omega_{1\mu M}}^N  &     \gamma_5\phi~ {l_M}^N    \\
                                  & \\
             \gamma_5\phi ~{l_N}^M  &    \gamma^\mu~ {\Omega_{2\mu M}}^N
      \end{array}\right) ~~~,~~~~~N,M=1,\ldots,5
\end{equation}
with the euclidean {\it beins}\footnote{They are orthonormalized and their
explicit expression is $E^a
= \left(\begin{array}{cc}
             \gamma^a    &     0    \\
                0        &  \gamma^a
      \end{array}\right)
$ with $a=1,..,4$, and $ E^5 = \left(\begin{array}{cc}
               0         &  \gamma_5 \\
            -\gamma_5    &     0
      \end{array}\right)
$.} such that $\Omega E^N = E^M \otimes {\Omega_M}^N$, and
${\Omega_N}^{M*}={\Omega_N}^M$. The ${l_M}^N$ are auxiliary fields.
Requiring compatibility with the canonical Riemaniann structure induced on
the space of one-forms by the spectral triple $({\cal A},{\cal H},D)$, and
null torsion condition
\bea
- {\Omega_P}^{N} \delta^{PM} + \delta^{NP} {\Omega_P}^M &=&0~~~,
\nonumber \\
dE^N-E^M{\Omega^N}_M&=&0~~~,
\eea
with $\delta^{NM}$ the five-dimensional Kronecker delta, one gets the
expression of ${\Omega_M}^N$ in terms of $\omega_\mu, \phi$ and $l$. In
particular ${\Omega_{1\mu M}}^N={\Omega_{2\mu M}}^N$ for $N,M =1,...4$, and
it corresponds to the Levi-Civita connection for the metric $g_{\mu \nu}=
e^a_\mu e^a_\nu$. It is now possible to calculate the Riemann tensor
${R_N}^M=d {\Omega_N}^M + {\Omega_M}^P{\Omega_P}^N$ and then the
Einstein--Hilbert action. Eliminating the auxiliary fields and continuing
to Lorentzian signature one gets
\begin{equation}
S_{EH} = \int_{\cal M}
\sqrt{-g}~\left({m_{Pl}^2 \over 16 \pi} {\cal R}-2\Lambda +{ 1 \over 2}
\partial_\mu\sigma\partial^\mu\sigma
\right)~d^4x~~~,
\end{equation}
where ${\cal R}$ stands for the scalar curvature of ${\cal M}$, and
$\sigma$ is a real scalar field such that $\phi = m_{Pl} \exp(-k\sigma)$,
with $k \equiv  \sqrt{4 \pi}/m_{Pl}$. The presence of the Planck mass in
the definition for $\sigma$ is due to the requirement of having $m_{Pl}$ as
the only mass scale in the model. Notice that this definition of $\sigma$
is the one giving a canonical kinetic term.

Using the Dirac operator with both gauge and gravitational connections we
combine the previous results to get a simple model describing interacting
gravity and gauge fields. The corresponding bosonic action is
\begin{eqnarray}
S_{B}  & = &\int_{\cal M}\sqrt{-g} \left\{
- { 1 \over 4} F_{\mu \nu} F^{\mu \nu}
- {\lambda \over 4!} \left[ |\varphi|^2 -
m^2 \exp\left(- 2 k \sigma\right)\right]^2    \right.
\nonumber\\
&+& \left. \left[{\partial}^{\mu} \varphi
{\partial}_{\mu} \varphi^* +  k
\partial _{\mu} |\varphi|^2 ~\partial^{\mu} \sigma +
\left({1\over2} + k^2 |\varphi|^2 \right)
\partial_{\mu} \sigma \partial^{\mu} \sigma \right]\right.
\nonumber\\
&+&
\left.{m_{Pl}^2 \over 16 \pi} {\cal R}-2\Lambda
\right\}~d^4x~~~,
\label{lagrab1}
\end{eqnarray}
where $F^{\mu \nu}$ is the usual $U(1)$ curvature tensor field, $\lambda$
is a positive coupling and we have redefined $\Phi
\equiv
\varphi~ g_F \left[n_{g} /  \tr(K^{\dag}K)
\right]^{1/2}$, and finally $m_{Pl} \equiv m~ g_F \left[n_{g} / \tr(K^{\dag}K)
\right]^{1/2}$, $g_F$ being the $U(1)$ gauge coupling.

We notice that the non diagonal elements of the matrix algebra, and of the
Dirac operator, have a twofold interpretation. On one side they are the
Higgs field in the gauge setting, whose natural scale is the electroweak
scale. On the other side they appear as the discrete component of the
Levi--Civita connection, with a natural gravitational (Planck) scale. It is
this dual role which solves the hierarchy problem in this setting. From Eq.
(\ref{lagrab1}) we see that the tree-level potential for the $\varphi$ and
$\sigma$ fields $V(\sigma,\varphi)$ takes the expression
\begin{equation}
V(\sigma,\varphi) = V_0 + {\lambda \over 4!} \left[ |\varphi|^2 -
m^2 \exp\left(- 2 k \sigma\right)\right]^2~~~.
\label{treepot}
\end{equation}
This potential is the same found in equation (19) of \cite{RS}. The
expression of the vacuum expectation value of $\varphi$ in terms of the
exponential of $\sigma$, allows to reduce its  natural scale $m \sim
m_{Pl}$ of several orders of magnitude with not fine tuned values for $k
\sigma$. This represents the natural solution of the hierarchy problem as
also invoked in RS.

It is interesting therefore to make an attempt to relate these two models,
and see to which extent the fact that the same equation \eqn{treepot} is
obtained is a coincidence, or it has some deeper meaning. The RS model is
based on the idea of a $4+d$ dimension spacetime with a nonfactorizable
geometry. The spacetime has a boundary composed of four dimensional branes.
For the simplest case of $d=1$ and two branes, the metric is
\be
ds^2=e^{-2kr_cx_5}\eta^{\mu\nu}dx_\mu dx_\nu+r_c^2dx_5^2~~~,
\ee
with $\eta^{\mu\nu}$ the ordinary 4-dimensional Minkowski metric,
$x_5\in[0,\pi]$, and all matter fields are constrained to live on the two
4-dimensional branes whose distance is $r_c$.

The two branes are very different, at $x_5=\pi$ (the visible brane) live
the fields which compose our world, while the fields living at $x_5=0$ are
basically unobservable. The RS potential for the Higgs field, as stated
above, coincides with \eqn{treepot} with the quantity $\sigma$ replaced by
the distance between the branes.

A deeper connection between the Randall--Sundrum model, branes and the
Connes--Lott scheme could emerge in the framework of the noncommutative
geometry of strings and branes. In string theory, what we call spacetime
comes from the low energy limit of a two dimensional conformal field
theory. Strings (open and closed) are described by a set of Vertex
Operators and their algebra\footnote{More precisely their $C^*$-algebra
completion.} form the {\em noncommutative geometry of strings} \cite{FG}.
Thus, the algebra of continuous functions on spacetime, which contains all
the information about the manifold, results to be the subalgebra of the
vertex operator algebra composed only of tachyon vertex operators after a
suitable projection \cite{LS}. In the case of open strings in the presence
of branes one has to consider the vertex operators at the endpoints. The
bosonic vertex operators for the emission of strings at the two endpoints,
will generate an algebra which is two copies of the algebra of function on
spacetime (the brane). This could be identified with the algebra
\eqn{alg2points}. From this point of view the two copies of spacetime are
nothing but the positions of the two endpoints of the strings on the brane.
These are the operators which lead to the noncommutativity of spacetime in
the presence of the antisymmetric tensor $B$ in the limit $\alpha'\to 0$
\cite{SW}, though in the RS and Connes--Lott the two sectors of spacetime
are still commutative. It is tempting to speculate that a generalization of
these models taking noncommutativity into full account can better describe
fundamental interactions\footnote{A first attempt to introduce the
noncommutativity of branes in the RS model has been made in
\cite{Ardalan}.}.

The algebra of the vertex operators at the two ends of the string is the
sum of two (commuting) subalgebras, each of them representing a copy of
spacetime. The same can be said for the Hilbert space. The details of the
particular string theory considered do not really matter.\\ From the
noncommutative geometry point of view, in the very low energy limit we thus
have two copies of spacetime, and a double Hilbert space. Closed strings
represent the bulk of spacetime and, as well known, their vertex operators
are responsible for gravitational interaction. They couple to both left and
right vertex operators. A full description of the interactions of open and
closed vertex operators (even at low energy) would require details of the
string theory, and probably a very heavy mathematical machinery. However,
as far the gravitational interaction along the brane is concerned, we
already know that if the string theory is related to general relativity, it
is described by a spin connection. As for the bulk, if we make an
approximation of retaining only the zero mode of the complicated vertex
operator interaction. Under this assumption we get a field interpolating
the two ends of the string (or the two copies of spacetime). This is the
Connes--Lott model.

To conclude we can say that probably the solution of the hierarchy problem
lies in the very structure of spacetime, and its investigation will give us
more fruitful surprises.

\subsubsection*{Acknowledgments}
We would like to thank G.~Landi, G.~Sparano and R.~Szabo for useful
discussions and correspondence.

\end{document}